# Gesture Recognition in Millimeter-Wave Radar Based on Spatio-Temporal Feature Sequences

Qun Fang[1] YiHui Yan[2] and GuoQing Ma[3]

[1] School of Computer and Information, Anhui Normal University, WuHui, China
fq0520@ahnu.edu.cn

[2] School of Computer and Information, Anhui Normal University, WuHui, China
yan_yihui@ahnu.edu.cn

[3] School of Computer and Information, Anhui Normal University, WuHui, China
2121012329@ahnu.edu.cn

## Abstract

*Gesture recognition is a pivotal technology in the realm of intelligent education, and millimeter-wave (mmWave) signals possess advantages such as high resolution and strong penetration capability. This paper introduces a highly accurate and robust gesture recognition method using mmWave radar. The method involves capturing the raw signals of hand movements with the mmWave radar module and preprocessing the received radar signals, including Fourier transformation, distance compression, Doppler processing, and noise reduction through moving target indication (MTI). The preprocessed signals are then fed into the Convolutional Neural Network-Time Domain Convolutional Network (CNN-TCN) model to extract spatio-temporal features, with recognition performance evaluated through classification. Experimental results demonstrate that this method achieves an accuracy rate of 98.2% in domain-specific recognition and maintains a consistently high recognition rate across different neural networks, showcasing exceptional recognition performance and robustness.*

## Keywords

## 1 Introduction

With the continuous advancement of the Internet of Things in Artificial Intelligence (AIOT), human-machine interaction has become increasingly crucial. Gesture interaction, due to its natural and efficient characteristics, has emerged as a hot research topic. It has found widespread applications across various domains, including autonomous driving[1-2] and smart home devices[3-4]. Users can engage in touchless interaction with digital devices through gestures, thereby enhancing the user experience.

There are various sensors capable of achieving touchless gesture interaction with digital devices, including cameras [5], WiFi [6], and millimeter-wave radar[7-9]. Although camera-based gesture recognition offers excellent recognition performance, practical application scenarios still pose significant challenges due to lighting conditions and privacy concerns. WiFi-based gesture recognition methods, constrained by their wavelength, struggle to recognize fine-grained gestures, limiting their practical usability. In contrast, millimeter-wave gesture interaction technology offers advantages such as high precision, strong robustness, privacy protection, and low power consumption.

To harness these advantages, extensive research has been conducted on gesture recognition using millimeter-wave radar, often combining traditional machine learning techniques with deep learning approaches. For instance, Zhang et al.[10] utilized Support Vector Machines (SVM) to classify micro-Doppler information, achieving an impressive accuracy of 88.56% in classifying four gestures at a distance of 0.3m. However, their method involved a complex manual feature extraction process, resulting in limited recognition accuracy. Dekker et al.[11] attempted to use 3D-CNN to classify three gestures, and the results indicated a recognition rate of 91%. However, 3D-CNN has limitations in terms of data resolution sensitivity and data requirements. Another study by Ref et al. [12] introduced a customized multi-branch Convolutional Neural Network (CNN) to automatically extract motion features from continuous gestures, achieving an accuracy of 95% in gesture classification. Nevertheless, the use of a single convolutional kernel in their approach limited its ability to fully capture and integrate temporal and spatial information of gestures. To overcome these limitations, Chen et al. [13]employed a CNN-Long Short-Term Memory (LSTM) architecture to capture both temporal and spatial information, effectively enhancing gesture recognition. However, CNN-LSTM models often require significant memory usage, possess high computational complexity, and are highly dependent on environmental factors.

In order to tackle these challenges and improve recognition accuracy and robustness, this paper introduces a gesture recognition method based on neural networks. We employ millimeter-wave radar to capture the raw signals of gesture movements and, subsequently, through preprocessing and neural network techniques, we can capture both temporal and spatial variations while reducing noise interference. This leads to increased accuracy and robustness in gesture recognition. Experimental results confirm the effectiveness of our proposed method, demonstrating its potential across various gesture recognition applications.

## 2 FMCW radar principle

### 2.1 Signal Model

The experiment utilizes the IWR1642, a commercially available low-cost MIMO radar module manufactured by Texas Instruments. This radar system is equipped with 2 transmitting antennas and 4 receiving antennas arranged horizontally. To achieve an equivalent configuration of an 8-antenna Uniform Linear Array (ULA), the radar employs Time Division Multiplexing (TDM) mode.

In the FMCW radar system, demodulation techniques [14]are commonly utilized. The received echo signal is mixed with the transmitted signal and then passed through a low-pass filter to extract the intermediate frequency (IF) signal. The IF signal model for a single scattering point can be represented as a sawtooth wave emitted by the FMCW radar [15]. The received and transmitted signals are fed into a mixer and subsequently filtered by a low-pass filter to obtain the IF signal. After undergoing I/Q sampling, the IF signal is converted into a discrete sequence of samples. The processing flowchart for the Frequency Modulated Continuous Wave (FMCW) radar is depicted in Figure 1.

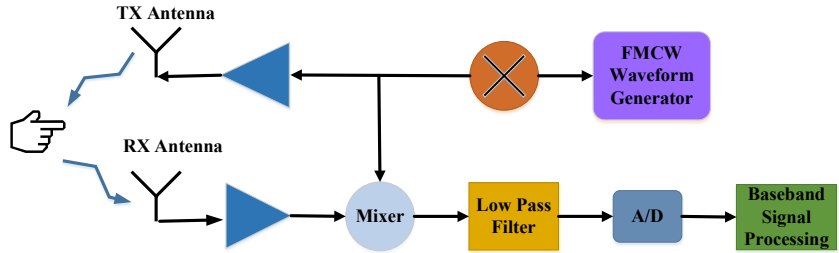

Fig．1：The FMCW radar processing flowchart

The radar transmit signal and echo signal are both characterized by a sawtooth waveform, and their interrelation is illustrated in Figure 2. Specifically, the transmit signal and echo signal for a single transmit cycle can be described as follows:

$$S_{TX}(t) = A_{TX} \cos\left(2\pi f_c t + \pi k t^2\right) \quad (1)$$

$$S_{RX}(t) = A_{RX} \cos\left(2\pi f_c (t-t_d) + \pi k (t-t_d)^2\right) \quad (2)$$

The parameter $k = B/T_c$ denotes the frequency modulation slope, where Tc represents the signal period, B stands for the signal bandwidth, fc represents the carrier frequency, $A_{TX}$ corresponds to the amplitude of the transmit signal, and $A_{RX}$ represents the amplitude of the echo signal. The echo signal undergoes mixing with a mixer and subsequent filtering through a low-pass filter to obtain an intermediate frequency signal.

$$S_{IF}(t) = \frac{1}{2} A_{TX} A_{RX} \cos\left(2\pi k \left(t_d t - \frac{1}{2} t_d^2\right) + 2\pi f_c t_d\right) \quad (3)$$

As the value of td is extremely small, it can be safely neglected in practical measurement scenarios. Consequently, the frequency of the intermediate frequency signal can be approximated using the following expression:

$$f_{IF}(t) = k t_d = \frac{B}{T_c} \cdot \frac{2R}{c} \quad (4)$$

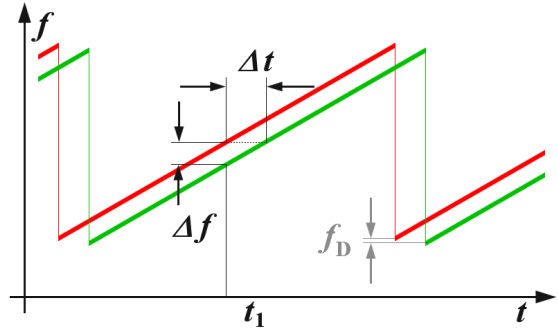

Fig．2 FMCW radar signal frequency versus time

## 2.2 Gesture signal pre-processing

This paper presents a radar signal preprocessing workflow designed for the efficient analysis and handling of received radar signals. This procedure involves conducting Fourier transformation, Doppler processing, noise reduction, and other techniques to transform the raw radar signals into clean Range-Frequency-Doppler Maps (RFDM). This preprocessing process facilitates subsequent feature extraction and classification.

### 2.2.1 Fourier transform

The radar's raw received signal is a time-domain signal, making it difficult to observe the signal's spectrum information. Fast Fourier Transform (FFT) is a rapid algorithm that converts time-domain signals into frequency-domain signals and efficiently calculates the Discrete Fourier Transform (DFT). In radar signal processing, the received echo signal is transformed from the time domain to the frequency domain. Fourier transform decomposes the signal into a series of composite forms of sine and cosine functions, allowing us to obtain the signal components at different frequencies and acquire spectrum information about the signal. By analyzing the spectrum distribution of the signal, we can extract better distance and velocity information of the target. Therefore, the conversion of the signal from the time domain to the frequency domain is a crucial step in radar signal processing, enabling us to better understand the signal characteristics and identify the target. The formula for Fast Fourier Transform is as follows:

$$X(k) = \sum_{n=0}^{N-1} x(n) e^{-j2\pi kn/N}, \quad k = 0, 1, \ldots, N-1 \quad (5)$$

Where x(n) is the time-domain signal, X(k) is the frequency-domain signal, and the value at the discrete frequency point *k* is obtained by the discrete Fourier transform of the time-domain signal *x(n)*.

### 2.2.2 Doppler processing

In radar signal processing, after range compression, the subsequent step involves acquiring the velocity information of the target. When a target is in motion, its echo signal undergoes a Doppler frequency shift. Therefore, the purpose of Doppler processing is to analyze this frequency shift and estimate the target's velocity.

Doppler processing is typically implemented using an FFT-based method known as Fast Time-Frequency Analysis (FTFA). In the FTFA approach, the range-compressed signal is initially transformed from the time domain to the frequency domain using a fast Fourier transform. Subsequently, the Doppler transform is applied to the frequency-domain signal of each range bin, enabling the extraction of the target's velocity information.

The objective of Doppler processing is to correct the echo signal on the frequency axis, effectively compensating for the Doppler frequency shift caused by the target's motion. This correction is achieved by multiplying the signal by a phase factor calculated based on the target's velocity. By employing this approach, the raw echo signal is recovered, facilitating subsequent target identification and tracking tasks.

### 2.2.3 MTI processing

In radar detection, the echo signals from clutter objects often exhibit stronger amplitudes compared to the target echo signals, leading to interference in target detection. To mitigate this interference and enhance radar sensitivity in target detection, the Moving Target Indication (MTI) signal processing technique is employed.

The MTI technique utilizes differencing operations to compare the echo data from multiple time instances. By subtracting the echo signals received at different time intervals, the MTI technique effectively suppresses the signals originating from stationary objects and mitigates the impact of clutter echo signals. As a result, the sensitivity of the radar system in detecting targets is improved.

Mathematically, the MTI processing can be represented by the following equation:

$$S_4(k,l,m) = S(k,l,m) - 4S(k,l-1,m) + 6S(k,l-2,m) - 4S(k,l-3,m) + S(k,l-4,m) \quad (6)$$

where $S(k, l, m)$ denotes the amplitude of the *k*th distance unit, the *l* th time unit, and the *m*th pulse-echo signal; $S_4(k, l, m)$ denotes the amplitude obtained after performing fourth-order MTI processing.

## 2.3 Neural Network Model

To effectively identify motion gestures, this study proposes a CNN-TCN-based spatiotemporal modeling approach for the modeling and classification of spatiotemporal data. The model divides the input RFDM into two parts: spatial features and temporal features. Specifically, the model consists of two components:

1. Frame Model: The CNN component extracts spatial features from each frame of the RFDM.

2. Sequence Model: The TCN component extracts temporal features from the RFDM's time series data.

The outputs of the frame model and sequence model are passed through fully connected layers and then merged for the final classification. This integration allows the CNN-TCN model to consider both spatial and temporal features simultaneously, resulting in improved classification accuracy. The network architecture of this model is depicted in Figure 3.

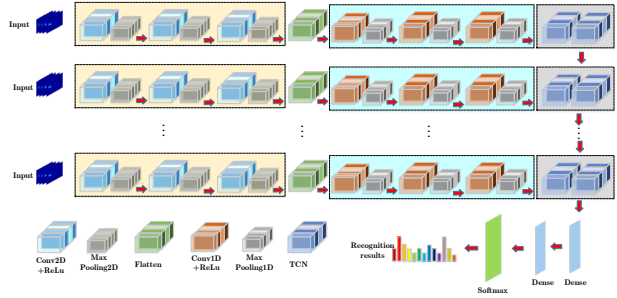

Fig．3：CNN-TCN neural network architecture

In particular, the framework model consists of three convolutional layers, a batch normalization layer, and two improved linear unit layers (LeakyReLU). These components are designed to extract spatial features from consecutive RFDM frames. The convolutional layers use kernels of different sizes to capture features at various scales. Subsequently, max-pooling layers are used to downsample the feature maps, reducing training time and improving the model's ability to generalize. Finally, one-dimensional CNNs are used to reduce the number of channels to 1/12 of the original, significantly reducing the model's complexity.

For consecutive frames, a TCN (Temporal Convolutional Network) framework is used to extract temporal features from the RFDM sequence. The proposed TCN differs from traditional structures as it adopts a streamlined design with flexible residual connections, as shown on the left side of Figure 4. Each TCN consists of three temporal blocks, as illustrated on the right side of Figure 4. These blocks include dilated convolution, causal convolution, LeakyReLU activation, and Dropout. Dilated convolution pads the input data, aligning the convolutional kernel with boundary pixels, while causal convolution exclusively uses past data to ensure that the output depends solely on the current and past inputs. This architecture effectively captures long-term dependencies within the sequence and allows for easy adjustment of the network's depth and width to accommodate different datasets and tasks. LeakyReLU activation mitigates the "neuron death" problem associated with traditional ReLU, improving model generalization and stability. Dropout reduces excessive interdependence among neurons, enhancing model generalization and reducing the risk of overfitting. In summary, this TCN structure effectively extracts temporal features from RFDM sequences while maintaining a low network complexity and training burden.

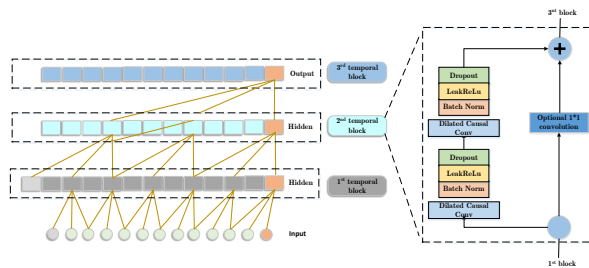

Fig. 4: The TCN network model diagram

## 3 Implemention

### 3.1 Dataset

Gesture recognition plays a crucial role in intelligent large-screen control, realizing contactless, natural, and intuitive interaction. However, gesture recognition encounters numerous challenges in various environments and positions, including lighting variations, background interference, and multi-path reflections. To tackle these challenges, we propose a deep learning-based gesture recognition network that leverages multi-dimensional features to enhance recognition accuracy and robustness. To demonstrate the performance and advantages of our proposed network, we meticulously design and curate a gesture dataset comprising seven distinct gestures performed in three different environments and five positions, as illustrated in the accompanying figure. Notably, our dataset exhibits distinctive characteristics when compared to existing gesture datasets:

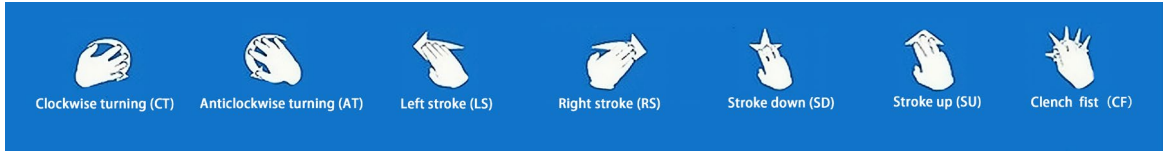

Figure 5: Data set gesture diagram

We have collected a total of 33,600 data samples for our dataset. A detailed description of the dataset is provided in Table 1.

Tab. 1: Detailed description of the dataset

| Environment | Data representation | Data volume |
|---|---|---|
| Classroom | 20Users×5Locations×(7Gestures×30Instances) | 21000Samples |
| Office | 10Users×3Locations×(7Gestures×30Instances) | 6300Samples |
| Conference hall | 10Users×3Locations×(7Gestures×30Instances) | 6300Samples |

## 4.2 Experimental equipment configuration

To collect gesture data, we utilized the IWR1642 radar system from Texas Instruments. This radar system is a millimeter-wave-based short-range solution that enables high-resolution gesture detection and recognition. It consists of two transmit antennas and four receive antennas, forming a two-dimensional array that provides spatial and Doppler information. Operating within the frequency range of 76-81 GHz, the radar system offers high bandwidth and sensitivity.

The radar system was positioned in front of the intelligent large-screen display, parallel to the microphone array, and connected to a computer running the gesture recognition network. To optimize data collection based on the experimental environment and gesture characteristics, we configured the radar acquisition board with specific parameters, as outlined in Table 2 below:

Table 2: IWR1642 acquisition board experimental parameters

| Parameter | Value | Parameter | Value |
|---|---|---|---|
| Number of range samples | 112 | Pulse Repetition Interval | 32.920us |
| Number of chirps | 128 | Frame time | 100.000ms |
| Sampling frequency | 6.250MHz | Max range | 104.095m |
| Carrier frequency | 77.144GHz | Max Dopple | ±29.512m/s |
| Bandwidth | 161.280MHz | Doppler resolution | 0.461m/s |

## 4.3 Neural Network Implementation

The experiment implemented the CNN-TCN gesture recognition neural network framework using the TensorFlow 2.0 framework and CPU: Intel I7-9750H. The network had an input feature map size of 32×414×1. The neural network consisted of a frame model and a sequence model. The frame model had three convolutional layers, where the number of kernels increased progressively (from 16 in the first layer to 64 in the last layer), but the convolution kernel size was fixed at 3×5. The output of the last Conv2D layer was flattened into a 1D vector and then fed into the sequence model. The sequence model had two enhanced TCNs and three fully connected layers to obtain gesture probabilities. The network was trained using the Adam optimizer, with a learning rate of 0.0005, a batch size of 128, and a training epoch of 100.

# 5 Evaluation

In this section, we first evaluate the accuracy of our neural network model in recognizing different gestures through data sets. We then evaluated the robustness of our model under different environments and factors, including new environments and new locations. Finally, we compared our model with other neural network methods to assess its performance.

## 5.1 Recognition results

In this study, we employed the Leave-One-Out Cross-Validation (LOOCV) method to evaluate the performance of our neural network in gesture recognition tasks. We divided the dataset into training and testing sets by selecting one individual's samples as the test set while using the remaining samples as the training set. This process was repeated for each individual in the same environment and location, allowing us to train and test the model comprehensively.

The results of the LOOCV were compiled into a confusion matrix, as illustrated in the Figure 6. The average recognition rate achieved was an impressive 98.4%. These findings demonstrate the high accuracy of our neural network in recognizing all the gestures studied. This success can be attributed to the similarity in features between the training and testing data.

Overall, these results affirm the effectiveness of our proposed neural network model in extracting relevant motion features for robust gesture recognition.

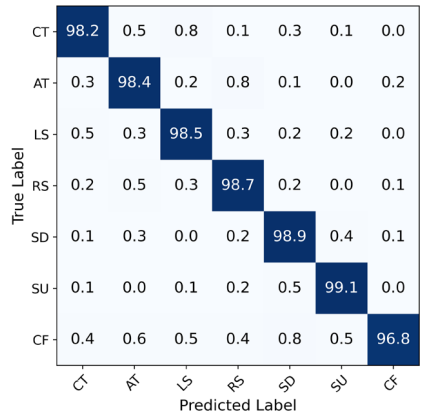

**Fig.6: Confusion Matrix for LOOCV Testing**

## 5.2 Location adaptability evaluation

We conducted experiments using data samples from different locations within the gesture dataset to validate the stability of our neural network model in recognizing gestures across various locations. The training set consisted of data from the (0.75, 0°) location, while the remaining locations were used for testing.

The results shown in Figure 7 indicate that the accuracy of gesture recognition was primarily affected by the spatial relationship between the hand and the radar sensor. When the sensor was too close to the hand, multipath effects led to a slight decrease in accuracy. Similarly, increasing the distance between the hand and the sensor resulted in a slight drop in accuracy due to improved signal-to-noise ratio. Additionally, the angle between the gesture and the radar sensor affected factors such as the main lobe range, received signal energy, and Doppler frequency.

Despite these factors, our model achieved consistently high accuracy across different locations, with an average recognition rate of 97.2%. This demonstrates the robustness of our model and its ability to adapt to variations in location that may impact gesture recognition.

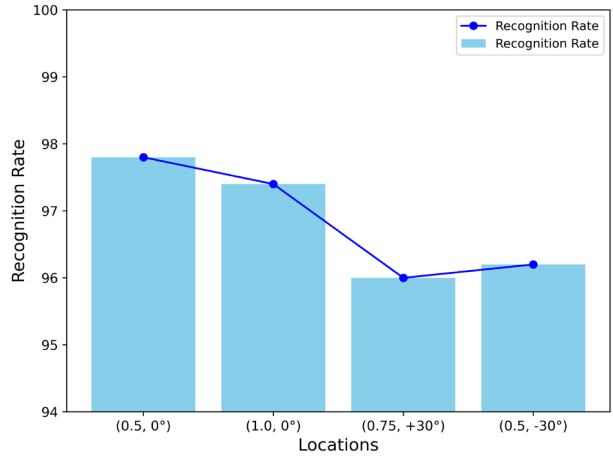

Fig.7: Accuracy of new location test.

## 5.3 Environmental adaptability evaluation

To validate the robustness of our neural network model in different environments, we considered variations in room sizes, placement of office equipment, and overall layout, which can result in different multipath effects. We collected gesture datasets from three different environments: a classroom dataset for training and separate datasets from a conference room and a lobby for testing.

The experimental results, as shown in the Figure 8, confirmed that different multipath effects indeed influenced the recognition performance of the neural network. However, despite these variations, the model exhibited high recognition accuracy in both environments. The recognition rate in the conference room was 97.0%, while in the lobby, it reached 98.5%. These high accuracy rates indicate the strong robustness of the model.

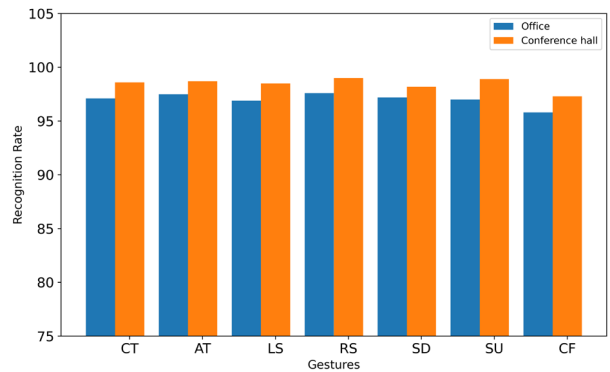

Fig.8: Accuracy of new environment test.

## 5.4 Neural network performance comparison

To further validate the effectiveness of the proposed network, we compared it with four other deep learning networks commonly used for action recognition: CNN, 3D-CNN, CNN-LSTM, and CNN-GRU. We evaluated their performance on our collected gesture dataset, and the results are summarized in Table 3.

Table 3: Recognition rate of different neural network model systems

| Serial number | Neural network model | Recognition accuracy (%) |
| --- | --- | --- |

| 1 | CNN | 80.6 |
| 2 | 3D-CNN | 87.7 |
| 3 | CNN-LSTM | 93.3 |
| 4 | CNN-GRU | 90.8 |
| 5 | CNN-TCN | 98.4 |

The CNN-TCN network outperformed the other networks in terms of recognition rates for each action category, demonstrating its superior ability to extract both spatial and temporal features. While CNN could only capture spatial features, 3D-CNN suffered from a large parameter count and limited flexibility in handling the temporal dimension. CNN-LSTM and CNN-GRU networks, although incorporating recurrent layers for temporal feature extraction, had higher computational complexity and limited modeling capabilities for long-term dependencies.

In contrast, the CNN-TCN network employed multi-scale spatio-temporal convolutional layers and fusion layers, allowing it to adaptively extract features at different scales and stages. It also dynamically fused information from multiple branches, resulting in superior performance across different action recognition tasks.

## 6 Conclusion

In this study, we utilized a millimeter-wave radar module to capture raw signals of hand gestures. By combining preprocessing techniques and convolutional neural network models, we successfully extracted spatio-temporal features and developed a contactless gesture recognition method based on millimeter-wave radar.The experimental results demonstrated that our method achieved high accuracy and robustness, overcoming the limitations of traditional gesture recognition techniques. It provides strong support for the advancement of smart education. However, there are still some limitations that need to be addressed. These include the high cost of radar signal acquisition equipment, limited scale and diversity of the dataset, and the need to improve the generalization ability of the classification model.

**FANG Qun**,

Born in 1972，Ph.D，professor.

His main research interests include signal processing, intelligent internet of things, computer vision and deep learning.

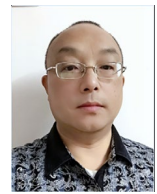

**YiHui Yan**

Born in 1997, postgraduate

His main research interests include radar signal processing, multi-sensor fusion, deep learning, and autonomous driving

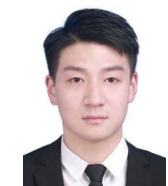

**GuoQing Ma**

Born in 2000, postgraduate

His main research interests include radar signal processing, behavioral perception, and deep learning

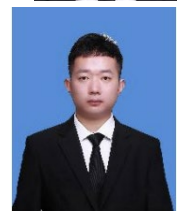